\documentstyle[12pt]{article}
\let\LARGE=\Large
\let\Large=\large
\let\large=\normalsize
\newcommand{\be}[3]{\begin{equation}  \label{#1#2#3}}     
\newcommand{\ee}{ \end{equation}}
\newcommand{\ba}{\begin{array}}
\newcommand{\ea}{\end{array}}

\def\beq{\begin{equation}}
\def\eeq{\end{equation}}
\def\beqa{\begin{eqnarray}}
\def\eeqa{\end{eqnarray}}

\def\fund{  \> {\vcenter  {\vbox
              {\hrule height.6pt
               \hbox {\vrule width.6pt  height5pt
                      \kern5pt
                      \vrule width.6pt  height5pt }
               \hrule height.6pt}
                         }
                   }
           \>\> }
\def\antifund{  \> \overline{ {\vcenter  {\vbox
              {\hrule height.6pt
               \hbox {\vrule width.6pt  height5pt
                      \kern5pt
                      \vrule width.6pt  height5pt }
               \hrule height.6pt}
                         }
                   } }
           \>\> }

\def\sym{  \> {\vcenter  {\vbox
              {\hrule height.6pt
               \hbox {\vrule width.6pt  height5pt
                      \kern5pt
                      \vrule width.6pt  height5pt
                      \kern5pt
                      \vrule width.6pt height5pt}
               \hrule height.6pt}
                         }
              }
           \>\> }

\def\symbar{  \> \overline{ {\vcenter  {\vbox
              {\hrule height.6pt
               \hbox {\vrule width.6pt  height5pt
                      \kern5pt
                      \vrule width.6pt  height5pt
                      \kern5pt
                      \vrule width.6pt height5pt}
               \hrule height.6pt}
                         }
              }
           } \>\> }
\def\anti{ \> {\vcenter  {\vbox
              {\hrule height.6pt
               \hbox {\vrule width.6pt  height5pt
                      \kern5pt
                      \vrule width.6pt  height5pt }
               \hrule height.6pt
               \hbox {\vrule width.6pt  height5pt
                      \kern5pt
                      \vrule width.6pt  height5pt }
               \hrule height.6pt}
                         }
              }
           \>\> }

\begin{document}
\thispagestyle{empty}
\rightline{HUB-EP-98/24}
\rightline{hep-th/9804074}
\vspace{2truecm}
\centerline{\bf \LARGE Superconformal $N=1$ Gauge Theories, 
$\beta$-Function
}

\centerline{\bf \LARGE Invariants  
and their Behavior under
Seiberg Duality
} 
\vspace{1.2truecm}
\centerline{\bf 
Andreas Karch\footnote{karch@qft1.physik.hu-berlin.de}\ , \ 
Dieter L\"ust\footnote{luest@qft1.physik.hu-berlin.de}\ and \
George Zoupanos\footnote{george.zoupanos@cern.ch}}

\vspace{.5truecm}
{\em 
\centerline{Humboldt-Universit\"at, Institut f\"ur Physik,
D-10115 Berlin, Germany}}

\vspace{1.0truecm}
\begin{abstract}
In this paper we 
discuss some aspects of the 
behavior of superconformal $N=1$ models under Seiberg's
duality. Our claim is that if an electric gauge theory is superconformal
on some marginal subspace  of all coupling constants then its magnetic
dual must be also superconformal on a corresponding moduli space
of dual couplings. However this does not imply that the magnetic dual of
a completely finite $N=1$ gauge theory is again finite.
Moreover
we generalize this statement 
conjecturing that also
for non-superconformal $N=1$ models the determinant of the
$\beta$-function equations is invariant under Seiberg duality.
During the course of this investigation we construct some superconformal
$N=1$ gauge theories which were not yet discussed before.
\end{abstract}

\bigskip \bigskip
\newpage

Seiberg \cite{seiberg} has conjectured that two different,
dual $N=1$ supersymmetric gauge theories lead to the same physics in the
infrared, where the strong coupling region of one gauge theory corresponds
to the weak coupling region of the other, and vice versa.
For example, $N=1$ SUSY QCD with gauge group $SU(N_c)$ and
$N_F$ quark flavors $Q_i$, $\bar Q_i$ ($i=1,\dots , N_F$) in the fundamental
respectively anti-fundamental representations of the gauge group
is dual to the $SU(N_f-N_c)$ SUSY QCD again with $N_F$ fundamental
quarks $q_i$, $\bar q_i$ plus  gauge singlets $M_{ij}$.
Many other examples, based on $SO(N_c)$ and $SP(N_c)$ gauge groups are
discussed in the literature.
To claim that two $N=1$ gauge theories are dual in above mentioned sense,
certain physical requirements have to be satisfied. Most notably these
are

\noindent (i) Agreement of global symmetries and global anomalies for dual
pairs.

\noindent (ii) Same gauge invariant operators (baryons, mesons).

\noindent (iii) Same behavior under deformations of the theory (turning
on/off vevs or mass terms).

In this note we shall discuss a new constraint which two dual models
should fullfil, and which to our knowledge has not been discussed
before in the literature.
To be more specific, we will discuss the all loop-order gauge plus
Yukawa
$\beta$-functions of $N=1$ gauge theories and argue that a certain
$\beta$-function determinant, which is build from the coefficients of the
$\beta$-function equations, is an invariant under Seiberg's duality,
i.e. should be the same for dual pairs.
Therefore the new ingredient emerging from the present
discussion is that even though the
$\beta$-functions change under Seiberg's duality, its determinant is unchanged.

This discussion is particularly relevant for $N=1$ models which are
completely finite (like $N=4$ gauge theories), with vanishing
$\beta$-functions as well as vanishing anomalous dimensions.
A bigger class of models is given by those which
contain at least one or several exactly marginal operators, 
which were first discussed
by Leigh and Strassler \cite{leigh}.
(Of course the models with marginal operators include
the completely finite models.)
The existence
of an  exactly marginal operator corresponds to having an arbitrary coupling
in the theory, called modulus.
On the fixed line of marginal couplings the theory is
superconformal, i.e. all $\beta$-functions are vanishing.
Following the work of Leigh and Strassler \cite{leigh}, 
a simple criterion for the existence of exactly marginal
operators is given by the vanishing of
the $\beta$-function determinant, which we mentioned above.

For dual models, the dimensions of the moduli space should agree, since
in the infrared there should be the same number of marginal operators
which can be used to deform the theory. Hence,
having found a marginal operator in one $N=1$ gauge theory,
there must be also the corresponding marginal operator in the dual
gauge theory. Therefore the $\beta$-function determinant must vanish
in the dual theory,
if it was zero in the
original model.
In other words, the dual magnetic model must be
also superconformal on the line
of fixed points if the electric
model is superconformal. More generally, we will argue that even in case
of non-vanishing $\beta$-function determinants, they nevertheless should
be the same for dual models; this statement will be shown to be true
for several examples. We will also discuss that the $\beta$-function
determinants are invariant under adding massive fields to the theory or
under integrating out fields via their equation of motions.

In a recent paper \cite{finite} we have investigated the Seiberg duals of
several types of all loop finite $N=1$ gauge theories, including
the finite $SU(5)$ GUT models.\footnote{The finite $SU(5)$ GUT
models successfully predict, among others, the top quark mass \cite{kmz}.}
As a result of this investigation we have seen that the dual of
a finite model is in general non-finite \cite{leigh,finite}.
(We \cite{finite} found that for one particular 
finite model, namely $SO(10)$ with matter
fields in $N_f=8$ vector and $N_q=8$ spinor representations, 
the dual is also supposed to be finite.)
However, as we discussed above, the marginal operators should be
preserved under Seiberg's duality, and consequently, the dual
of a finite $N=1$ must have at least one marginal operator. 
In this way, non-finite $N=1$ gauge theories with marginal operators
may belong to the same universality class as finite $N=1$ models.
Both, the electric theory as well its magnetic dual are superconformal on the
moduli space of marginal couplings.

Recently, a class of all order finite $N=1$ models
were realized as gauge theories on  D3 branes with an orbifold
action on the space transversal to the branes \cite{KaSi}. 
In this context the
finiteness of these models follows from a duality to type IIB string theory
on $AdS^5\times S^5$ geometry, which should hold in the large $N$
limit of the gauge theory \cite{maldacena}. The $SO(4,2)$ symmetry of $AdS^5$ 
translates into the superconformal group of the gauge theory which
lives at the four-dimensional boundary of $AdS^5$.
We will analyze the $\beta$-function determinants and the Seiberg
duals for the type of $N=1$ models 
constructed in \cite{KaSi}.\footnote{For a construction of
these models by string compactification, see \cite{ibanez}.}
Also note that in the brane picture of Hanany and Witten \cite{HW},
which is $T$-dual to construction via D3 branes, it was
recently shown \cite{HSU}
that in finite theories with vanishing $\beta$-functions
the branes are not bent.

Consider a $N=1$ supersymmetric gauge theory based on the gauge group 
$G=\prod_{l=1}^N G_l$ with corresponding gauge couplings $g_l$
and $M$ chiral superfields $\phi_i$ 
$(i=1,\dots , M)$ in the 
$ R_{i,l}=( R_{i,1},R_{i,2},\dots ,
 R_{i,N})$
representation of the group $G$.
The, in general non-renormalizable, superpotential $W$ is of
the form 
\begin{equation}
W=\sum_{\alpha}h_\alpha\prod_i\phi_i^{n_{i,\alpha}},
\quad \alpha=1,\dots, L,
\end{equation}
where $n_{i,\alpha}$ is the number of superfields $\phi_i$ in each term of $W$.
Then the exact, i.e. all orders $\beta$-functions 
$\beta_m$, $m=1,\dots ,N,N+1,\dots ,N+L$, for the gauge couplings
$g_l$ and the Yukawa couplings $h_\alpha$ are given by the
following set of equations \cite{shifman}:
\begin{eqnarray}
\beta_{g_l}&=&-\biggl(3C_2(G_l)-\sum_iT(R_{i,l})
\biggr)-\sum_iT(R_{i,l})\gamma_i
=B_{l,0}+\sum_iB_{l,i}\gamma_i,\nonumber\\
\beta_{h_\alpha}&=&\biggl(-3+\sum_in_{i,\alpha}\biggr)+
{1\over 2}\sum_in_{i,\alpha}\gamma_i=
B_{N+\alpha ,0}+\sum_iB_{N+\alpha, i}\gamma_i.\label{betfct}
\end{eqnarray}
Here,  $\gamma_i$ is the anomalous dimension of the
superfield $\phi_i$, $C_2(G_l)$ is the quadratic Casimir of the adjoint
representation of $G_l$, and $T(R_{i,l})$ is the index of
the representation $R_i$ with respect to $G_l$.
The anomalous dimensions are in general unknown functions of the
couplings constants $g_l$ and $h_\alpha$.

We see that the $\beta$-functions are given by a set of linear equations
in the anomalous dimensions $\gamma_i$ with a $\lbrack(N+L)\times
(M+1)\rbrack$-dimensional coefficient matrix
$B_{m,n}$,  $n=0,\dots ,M$.
Note that the number of columns can be smaller than $M+1$ in case
some of the superfields $\phi_i$ have the same anomalous dimensions; this
situation will apply if the superpotential is invariant under some
global symmetries. 
The matrix elements $B_{l,0}$, $l=1,\dots, N$, are nothing else
than the one loop $\beta$-function coefficients $\beta_l^{(1)}$ of the
gauge couplings. 
On the other hand,
the elements $B_{N+\alpha,0}$, $\alpha=1,\dots , L$, denote mass dimensions
of the coupling constants $h_\alpha$.

The condition for the existence of marginal operators \cite{leigh}
is that all $N+L$ $\beta$-function equations $\beta_m$ in eq.(\ref{betfct})
simultaneously vanish on some particular locus in the space of coupling 
constants
$(g_l,h_\alpha)$. The anomalous dimensions are functions of
the coupling constants, and therefore the 
vanishing of the $\beta$-functions puts
$N+L$ conditions on the $N+L$ couplings. Since we are not looking for
isolated fixed point solutions of the equations $\beta_m=0$
but for manifolds of fixed points, marginal operators are
present if the $\beta_m$
are linearly dependent functions in the anomalous dimensions
$\gamma_i$. More precisely, if we find that 
$r$ $\beta$-function equations are linearly dependent, then there is a
$r$-dimensional moduli space of marginal couplings 
$g^{(0)}_k$, $k=1,\dots , r$, which can be arbitrarily
chosen. On this manifold of fixed points the theory
is supposed to be superconformal. The $r$ equations
\begin{equation}
\sum_mc_{r,m}\beta_m=0
\end{equation}
with non-vanishing coefficients $c_{r,m}$
define the marginal couplings 
as functions of the couplings $g_l$ and 
$h_\alpha$,
but only implicitly, since we do not know the precise form of the all order
functions $\gamma_i(g_l,h_\alpha)$.
One might expect that an $S$-duality
group is acting on the marginal couplings
$g^{(0)}_k$, which relates strong and weak coupling in the manifold
of marginal operators.

For simplicity consider now the case that $B_{m,n}$ is a square
matrix of dimension $(N+L)\times (N+L)$, which means that we assume
that the number of different anomalous dimensions of
chiral superfields is one less than
the number of different couplings. Then the condition for the existence of
marginal operators can be simply stated as follows:\footnote{If $B$ is
not a square matrix this condition is generalized to
$\det(BB^T)=0$.}
\begin{equation}
\det ~B=0.
\end{equation}
The number of zero eigenvalues of $B$ coincides with the number $r$
of marginal operators.

Now consider all order finite $N=1$ gauge theories, which
are a subclass of the models with marginal operators. In the present
framework
$N=1$ gauge theories are finite if {\it all} $\beta$-functions $\beta_m$ 
{\it and
all} anomalous dimensions $\gamma_i$ vanish to all orders in perturbation
theory.
Then the $\beta$-functions eq.(\ref{betfct}) 
immediately imply that all constant terms
in these equations must be zero. In other words, finiteness requires that  all
one loop gauge $\beta$-functions $\beta_l^{(1)}=B_{l,0}$ vanish, and that all
classical couplings in the superpotential are dimensionless, i.e.
$B_{N+\alpha,0}=0$. 
If these conditions are satisfied the vanishing of the $\beta$-functions
provide an homogeneous system of linear equations in the anomalous
dimensions $\gamma_i$, which puts, as in the
case of  marginal operators, $N+L$
conditions on $N+L$ coupling constants. If $r$ of these linear 
equations are linearly dependent,
the theory is supposed to be finite on the $r$-dimensional
submanifold of free couplings $g^{(0)}_k$. The remaining couplings are not any
more independent, but are functions of the $g^{(0)}_k$. 
So as a criterion for finite $N=1$ theories we require that
\begin{equation}
\det ~ B=0,\quad {\rm with}\quad B_{l,0}=0,\quad B_{N+\alpha,0}=0.
\end{equation}
This condition is in agreement with a theorem \cite{LPS}\footnote{See also
\cite{kazakov}.}
which states that a $N=1$ gauge theory is all orders finite if
the one loop $\beta$-functions 
and the one loop anomalous dimensions
are vanishing and if there exist non-degenerate
solutions of the coupling constant reduction equations \cite{Zimmer}.
So for finite $N=1$ theories, the number of free coupling constants is
in general reduced, which just means that we are moving
in the $r$-dimensional subspace of marginal operators.
Also note that the vanishing of the one 
loop $\beta$-functions $\beta^{(1)}$ and
one loop anomalous dimensions $\gamma_i^{(1)}$ 
ensures the finiteness at the two loop level, too \cite{PW}. 

Now let us discuss the presence of marginal operators 
in the dual Seiberg picture. 
The dual model depends very much on the details of the original
gauge theory. Let us  denote  the dual gauge group by
$\tilde G=\prod_{l=1}^{\tilde N} \tilde G_l$
with matter fields $\tilde\phi_i$ ($i=1,\dots ,\tilde M$), 
couplings $\tilde g_l$ and
$\tilde h_\alpha$ ($\alpha=1,\dots, \tilde L$), anomalous
dimensions $\tilde\gamma_i$  and
$\beta$-functions $\tilde\beta_l$, $\tilde\beta_\alpha$ with
coefficient matrix $\tilde B$.
$N=1$ Seiberg duality states that two dual
models flow to the same infrared fixed points, which means that in the
infrared they are physically equivalent. So in the infrared, also the
dual theory must have the same marginal operators 
with coulings $\tilde g^{(0)}_k$ ($k=1,\dots , r$) as the original
model. However this requirement should not be restricted to the
infrared since the marginal couplings do not run. 
Therefore, for consistency one has
to demand that
\begin{equation}
\det~\tilde B=0
\end{equation}
in case $\det B=0$ in the original model.
However this requirement does not mean that the dual of a finite
$N=1$ theory is again a finite model, as it was already observed
in \cite{finite}. It only  means that the reduced
subspace of marginal couplings
must agree for dual pairs. 
Therefore making a statement
concerning invariants under Seiberg's duality it is not the notion of
finiteness but only the notion of being superconformal on the manifold of
fixed lines which remains invariant.
In the infrared, where the electric and the dual magnetic model describe the
same physics, there is no difference between a superconformal and a
completely finite $N=1$ gauge theory. Comparing the electric theory and
its magnetic dual,
 there should be a simple
relation which maps the marginal couplings in two
dual pairs onto each other:
\begin{equation}
\tilde g^{(0)}=\tilde g^{(0)}_k(g^{(0)}_k).
\end{equation}
However the couplings with non-vanishing
$\beta$-functions are not immediately related to each other, since
in the ultraviolet the two theories are not supposed to be identical.

Now we want to go one step further, conjecturing that the
$\beta$-function determinants $B$ and $\tilde B$ are in fact invariant
(up to a group theoretical factor) under Seiberg duality even in the case
they do not vanish, i.e. in the case the model is not superconformal and
the reduction of coupling constants is not possible.
We do not have a firm proof for this assertion, but we will show 
in the following that it
holds for many explicitly constructed dual pairs. 
The idea behind this conjecture is that  $\det B$ corresponds in some
sence to an overall coupling constant whose $\beta$-function
is not changed by the Seiberg duality. If the gauge theory can be derived
from string theory, this overall coupling might be directly
related to the string coupling constant.

As the first and simplest example consider 
supersymmetric QCD (SQCD) 
with $N_f$ flavors of quarks in the $(N_c+\bar N_c)$ representation 
of the gauge
group $G=SU(N_c)$ with
gauge coupling $g$. The corresponding chiral superfields will be denoted by
$Q_i$ and $\bar Q_i$, $i=1, \dots , N_f$.
For concreteness we assume that the superpotential contains only one
Yukawa coupling constant $h$ and consists of all 
possible $SU(N_f)$-flavor symmetric combinations of the 
baryons $B\sim Q^{N_c}$ ($N_f\geq N_c$):
\begin{equation}
W=h(Q_{i_1}Q_{i_2}\dots Q_{i_{N_c}}+\bar Q_{i_1}\bar Q_{i_2}\dots 
\bar Q_{i_{N_c}}+\dots ).
\end{equation}
Furthermore, the $SU(N_f)$ flavor symmetry  implies that all
anomalous dimensions are the same: $\gamma_i=\gamma$.
Then the corresponding two $\beta$-function equations lead to the following
matrix $B$:
\begin{equation}
B=\pmatrix{N_f-3N_c&-N_f\cr N_c-3&{N_c\over2}}.
\end{equation}
Its determinant is given by
\begin{equation}
\det~B={3\over 2}N_cN_f-{3\over 2}N_c^2-3N_f.
\end{equation}
The condition for the existence of a marginal operator, $\det B=0$, is
a diophantic equation in $N_c$ and $N_f$, and we found solutions 
for 

\noindent (i) $N_c=3$, $N_f=9$, 

\noindent (ii) $N_c=4$, $N_f=8$,

\noindent  (iii) $N_c=6$, $N_f=9$.

\noindent
The first case (i) corresponds to a finite model upon reduction to one
coupling constant; it is Seiberg dual to the non-finite case (iii).
The second solution (ii) is non-finite but self-dual under Seiberg
duality.

Let us  consider the Seiberg dual of SQCD in more detail. It is given
by the dual gauge group $\tilde G=SU(N_f-N_c)$ and $N_f$ fundamental
quarks $q_i$, $\bar q_i$ plus  gauge singlet mesons $M_{ij}$.
The dual superpotential is given by
\begin{equation}
\tilde W=M_{ij}q_i\bar q_j+\tilde h
(q_{i_1}q_{i_2}\dots q_{i_{N_f-N_c}}+\bar q_{i_1}\bar q_{i_2}\dots 
\bar q_{i_{N_f-N_c}}+\dots ).
\end{equation}
The dual $\beta$-functions now provide the following
matrix $\tilde B$ with respect to the two couplings $\tilde g$, 
$\tilde h$ and the anomalous dimension $\tilde\gamma$ of the fields $q_i$ and
$\bar q_i$:\footnote{One could also include the coupling constant of
term $M_{ij}q_i\bar q_j$ plus the anomalous dimension of the fields
$M_{ij}$ in the matrix $\tilde B$. However the determinant of the
corresponding $3\times 3$
matrix $\tilde B$ is only changed
by an irrelevant numerical factor compared to
eq.(\ref{dualbsun}). Since the coupling
of $M_{ij}q_i\bar q_j$ is  not independent anyway, we prefer not to include
this term (see the discussion in \cite{ohme}).}
\begin{equation}
\tilde B=\pmatrix{3N_c-2N_f&-N_f\cr N_f-N_c-3&{N_f-N_c\over2}}.\label{dualbsun}
\end{equation}
It is easy to show that
\begin{equation}
\det ~B=\det ~\tilde B,
\end{equation}
as advocated before.

Next let us discuss the $\beta$-function determinants and Seiberg duality
\cite{seiberg,seiintr}
for the gauge group $G=SO(N_c)$ with $N_f$ quarks $Q_i$ in the
fundamental vector representation. We discuss two types of superpotentials.
The first type is given, like for the $SU(N_c)$ case, by the
baryon superfield $B\sim Q^{N_c}$:
\begin{equation}
W=h(Q_{i_1}Q_{i_2}\dots Q_{i_{N_c}} 
+\dots ).
\end{equation}
Then we derive the following $\beta$-function matrix:
\begin{equation}
B=\pmatrix{N_f-3(N_c-2)&-N_f\cr N_c-3&{N_c\over2}}.
\end{equation}
Its determinant is given by
\begin{equation}
\det~B={3\over 2}N_cN_f-{3\over 2}N_c^2-3N_f+3N_c.\label{detso1}
\end{equation}
One marginal operator is present, if $N_c=N_f$, and the model is finite
if $N_c=N_f=3$.

The dual theory is based on the
gauge group  $\tilde G=SO(N_f-N_c+4)$ and contains $N_f$
fundamental quarks $q_i$ and the gauge singlet meson fields $M_{ij}$.
To obtain the dual superpotential it is important to remember that
the gauge invariant baryon operators $B\sim Q^{N_c}$ in the
electric theory are mapped to the following gauge invariant
operator on the magnetic side:
$B\rightarrow W_\alpha^2q^{N_f-N_c}$. Here, $W_\alpha$ is the chiral gauge
field strength superfield. 
Therefore the dual superpotential takes the following form:
\begin{equation}
\tilde W=M_{ij}q_i\bar q_j+\tilde h
(W_\alpha^2q_{i_1}q_{i_2}\dots q_{i_{N_f-N_c}} 
+\dots ).
\end{equation}
Since the field $W_\alpha$
has mass dimension ${3\over 2}$,
the dual $\beta$-functions are encoded in the following 
matrix $\tilde B$:
\begin{equation}
\tilde B=\pmatrix{3N_c-2N_f-6&-N_f\cr N_f-N_c&{N_f-N_c\over2}}.
\end{equation}
It is easy to show that again
the dual magnetic determinant agrees with the electric determinant
eq.(\ref{detso1}).

We can also start in the electric $SO(N_c)$ gauge theory with the
superpotential
\begin{equation}
W=h
(W_\alpha^2Q_{i_1}Q_{i_2}\dots Q_{i_{N_c-4}} 
+\dots ).
\end{equation}
Now the $\beta$-function matrix looks like
\begin{equation}
B=\pmatrix{N_f-3(N_c-2)&-N_f\cr N_c-4&{N_c-4\over2}},
\end{equation}
and the corresponding determinant is given by
\begin{equation}
\det~B={3\over 2}N_cN_f-{3\over 2}N_c^2-6N_f+9N_c-12.\label{detso}
\end{equation}
We see that with this superpotential, which includes the gauge field
strength $W_\alpha$, the gauge theory possesses a marginal  operator
for $N_c=4$ and $N_f$ arbitrary. The model is finite for $N_c=4$ and
$N_f=6$. 
On the dual magnetic side the superpotential is given in terms of the
baryonic operators:
\begin{equation}
\tilde W=M_{ij}q_i\bar q_j+\tilde h
(q_{i_1}q_{i_2}\dots q_{i_{N_f-N_c+4}} 
+\dots ).
\end{equation}
This leads to the matrix
\begin{equation}
\tilde B=\pmatrix{3N_c-2N_f-6&-N_f\cr N_f-N_c+1&{N_f-N_c+4\over2}},
\end{equation}
whose determinant agrees with eq.(\ref{detso}).

The next example we are considering is given by the product gauge
group $G=SU(N_c)\times SU(N_c)\times SU(N_c)$ with gauge couplings $g_1$,
$g_2$ and $g_3$. As matter fields we take $N_f$ fields in the representations
\begin{equation}
Q_1=(N_c,\bar N_c,1),\quad Q_2=(1,N_c,\bar N_c),\quad Q_3=(\bar N_c,1,N_c),
\end{equation}
with anomalous dimensions $\gamma_1$, $\gamma_2$, $\gamma_3$.
The cubic superpotential is given by
\begin{equation}
W=h(Q_1Q_2Q_3+\dots ).
\end{equation}
Computing the $\beta$-function equations for the four couplings
$g_1$, $g_2$, $g_3$ and $h$ we derive the following matrix $B$
(the first three rows correspond to the gauge $\beta$-functions,
the last three colums belong to the three $\gamma_i$):
\begin{equation}
B=\pmatrix{N_cN_f-3N_c&-{1\over 2}N_cN_f&0&-{1\over 2}N_cN_f\cr
N_cN_f-3N_c&-{1\over 2}N_cN_f&-{1\over 2}N_cN_f&0\cr
N_cN_f-3N_c&0&-{1\over 2}N_cN_f&-{1\over 2}N_cN_f\cr
0&{1\over 2}&{1\over 2}&{1\over 2}}.
\end{equation}
The corresponding determinant looks like
\begin{equation}
\det ~B={3\over 8}(N_f-3)N_c^3N_f^2.
\end{equation}
We see that $\det B$ vanishes for $N_f=3$. Then the model is completely finite.
The case $N_f=3$  can be obtained by $Z_3$-orbifoldizing an $N=4$, $SU(N_c)$
gauge theory and was recently discussed in the context of $AdS^5\times S^5$
geometry in type IIB superstrings \cite{KaSi}.

Now let us now construct the Seiberg dual of this model applying the
procedure outlined in \cite{schmaltz}.
As an abbreviation we introduce the following parameters:
$a=N_f-1$, $b=N_f^2-N_f-1$, $c=(N_f^2-N_f-1)(N_f-1)N_f-1$.
The dual gauge group is then given by
$\tilde G=SU(aN_c)\times SU(bN_c)\times SU(cN_c)$. The massless dual matter
fields transform under this gauge group as
\begin{eqnarray}
& & cN_f ~ \times ~\lbrack q_1=(aN_c,\overline{bN_c},1)\rbrack\nonumber\\
& &bN_f ~ \times ~\lbrack q_2=(\overline{aN_c},1,cN_c)\rbrack\nonumber\\
& &aN_f ~ \times ~ \lbrack q_3=(1,bN_c,\overline{cN_c})\rbrack.
\end{eqnarray} 
Note that in deriving this dual spectrum we used several mass terms between
mesons fields and quarks to decouple these states from the massless
spectrum.
The dual superpotential is given as
\begin{equation}
\tilde W= \tilde h(q_1q_2q_3+\dots ).
\end{equation}
The dual $\beta$-functions lead to the following matrix $\tilde B$:
\begin{equation}
\tilde B=\pmatrix{N_cN_fbc-3N_ca&-{1\over 2}N_cN_fbc&0&-{1\over 2}N_cN_fbc\cr
N_cN_fac-3N_cb&-{1\over 2}N_cN_fac&-{1\over 2}N_cN_fac&0\cr
N_cN_fab-3N_cc&0&-{1\over 2}N_cN_fab&-{1\over 2}N_cN_fab\cr
0&{1\over 2}&{1\over 2}&{1\over 2}}.
\end{equation}
Then the dual determinant is derived to be
\begin{equation}
\det ~\tilde B={3\over 8}(N_f-3)N_c^3N_f^2abc.
\end{equation}
Up to the group theoretical factor 
$(abc)$ the dual determinant $\det\tilde B$ agrees with $\det B$.

For $N_f=2$ the theory is selfdual. Note that in this case the
determinant does not vanish. This is in some way a very surprising
result, since it has been believed that selfduality
only appears in conection with marginal lines connecting the
two theories.

As a last example consider the model discussed in \cite{SO10}. The duality
relates $Spin(10)$ gauge theory with vectors and spinors to an
$SU \times Sp$ product gauge group with a symmetric tensor representation
for the $SU$ factor.

The electric theory has gauge group $G=Spin(10)$ with matter fields
transforming as
\begin{equation}
V= N_f \cdot 10, \quad S=N_q \cdot 16.
\end{equation}
The dual theory has $G=SU(\tilde{N_c}=N_f+2 N_q-7) \times Sp(2 N_q -2)$ gauge
group and matter fields transforming as
$$
q=N_f \cdot (\fund,1), \quad q'=2 \cdot (\fund,\fund), \quad
\bar{q}=(2N_q-1)\cdot  (\antifund,1), \quad s=(\overline{\mbox{sym.}},1),
\quad t=(2N_q-2) \cdot (1,\fund),
$$
\begin{equation}
M= (N_f+1)N_f/2 \cdot (1,1), \quad N= N_f (2N_q-1) \cdot (1,1)
\end{equation}

The dual theory has a superpotential given by
\begin{equation}
\tilde{W}= \lambda_1 M q s q + \lambda_2 N q \bar{q} +
\lambda_3 q' s q' +
\lambda_4 \bar{q} q' t
\end{equation}

In order to obtain a quadratic matrix $B$ at the
electric side we should consider a superpotential with
two coupling constants,
for example any of the following three
\begin{eqnarray}
 W_1 &=& h_1 S^4 + h_2 S^2 V^3 \\
 W_2 &=& h_1 S^2 V^5 + h_2 S^2 V^3 \\
 W_3 &=& h_1 S^4 + h_2 S^2 V.
\end{eqnarray}
According to the operator mapping of \cite{SO10} the corresponding
dual operators are
\begin{eqnarray}
 \tilde{W}_1 &=& \tilde{h_1} t^2 + \tilde{h_2} q^{N_f-3} q^{' 2N_q-4} \\
 \tilde{W}_2 &=& \tilde{h_1} q^{N_f-5} q^{' 2N_q-2}+ 
\tilde{h_2} q^{N_f-3} q^{' 2N_q-4} \\
 \tilde{W}_3 &=& \tilde{h_1} t^2 + \tilde{h_2} N.
\end{eqnarray}

Now it is easy to calculate the determinants on both sides, which
are determinants  of 3 by 3 matrices in the electric theory, while on the 
magnetic side we deal with 8 by 8 matrices.\footnote{Note
that we have included here the couplings $\lambda_1$ and
$\lambda_2$ of the meson fields $M$ and
$N$.} We obtain
\begin{eqnarray}
\det{B}_1 = 6 (-24 +2 N_f +3 N_q) & \quad& \det{\tilde{B}_1}= \frac{3}{16}
(24-2N_f-3N_q) \tilde{N_c} \\
\det{B}_2 = 6(8 - N_f) & \quad& \det{\tilde{B}_2}= \frac{3}{16}
(N_f-8) \tilde{N_c} \\
\det{B}_3 = 6(8-N_q)  & \quad& \det{\tilde{B}_3}= \frac{3}{16}
(N_q-8) \tilde{N_c} 
\end{eqnarray}
Note that the models are superconformal, but not completely 
finite for those values of
$N_f$ and $N_q$ which 
lead to vanishing determinants $\det B_{1,2,3}$.\footnote{One can obtain
a completely finite $SO(10)$ model with $N_f=N_q=8$ adding the superpotential
$W\sim SSV$ on the electric side. After some symmetry breaking also the
magnetic dual of this model is supposed to be completely finite
\cite{finite}.}


Another interesting case to consider is the decoupling of massive
fields. Consider an arbitrary gauge theory with 2 kinds
of fields, which we call  $Q$ and $A$.
The superpotential
is given by $W=h Q A^p + m Q^2$.
We now will show that the determinat is the same (up to
a factor of 2) whether we keep the
massive quark and the mass term or integrate it out, as it should
be. The massive quark cannot affect the IR behaviour. This
is another generalization of the results of Leigh and Strassler,
who allready found that the existence of marginal operators (that
is the zeros of the determinant) remain under integrating
out massive fields.

Before integrating out the fields we have three couplings, the gauge
coupling, $h$ and $m$. The $B$ matrix is hence given as
\begin{equation}
B=\pmatrix{C-\mu_Q - \mu_A& \mu_Q& \mu_A
\cr p-2 & 1/2 & p/2 \cr -1 &1 &0}
\end{equation}
where C is the contribution of the gauge fields to the 1-loop
$\beta$ function and the $\mu_{Q,A}$ denote the indeces of the
representations of $Q$ and $A$.
With this one obtains
\begin{equation}
\det{B} = -\frac{p}{2} C +\frac{3}{2} \mu_Q (p-1).
\end{equation}
After integrating out we are just left with the
field $A$ and an effective superpotential $W= - \frac{h}{4m} A^{2p}$,
yielding the reduced matrix
\begin{equation}
B=\pmatrix{C- \mu_A & \mu_a \cr 2p-3 & p}
\end{equation}
and hence
\begin{equation}
\det{B} = -p C +3 \mu_Q (p-1).
\end{equation}

\vskip1cm
\noindent{\bf Acknowledgements:}

\noindent G.Z. is on leave from the
  Physics Department,
 National Technical University,
 Zografou Campus,
 GR-15780 Athens, Greece.
 Work partially supported by the E.C. projects
FMBI-CT96-1212; ERBFMRXCT960090 and the Greek projects,
PENED95/1170;1981. The work of A.K. and D.L. is supported by the DFG.

\end{document}